%% file: ANA-EXOT-2020-27-PAPER.tex
\documentclass[PAPER, atlasdraft=false, texlive=2023, UKenglish, cernpreprint, texmf, orcidlogo]{atlasdoc}
 
\usepackage[subcaption]{atlaspackage}
\usepackage{atlasbiblatex}
 
\usepackage{atlasphysics}
 
\ifthenelse{\boolean{AtlasDraft}}{
}{}
 
\graphicspath{{logos/}{figures/}}
 
\usepackage{ANA-EXOT-2020-27-PAPER-defs}

\AtlasTitle{Combination of searches for pair-produced leptoquarks
at $\sqrt{s} = 13\,$TeV with the ATLAS detector}

\AtlasAbstract{
A statistical combination of various searches for pair-produced leptoquarks is presented, using
the full LHC Run~2 (2015--2018) data set of $139\,$fb$^{-1}$ collected with the ATLAS detector from proton--proton collisions at a centre-of-mass energy of $\sqrt{s}=13\,$TeV. All possible decays of the leptoquarks into quarks of the third generation and charged or neutral leptons of any generation are investigated. Since no significant deviations from the Standard Model expectation are observed in any of the individual analyses, combined exclusion limits are set on the production cross-sections for scalar and vector leptoquarks. The resulting lower bounds on leptoquark masses exceed those from the individual analyses by up to 100\,GeV, depending on the signal hypothesis.
}

\AtlasRefCode{EXOT-2020-27}
 
\PreprintIdNumber{CERN-EP-2023-288}

\AtlasJournalRef{\PLB 854 (2024) 138736}
\AtlasDOI{10.1016/j.physletb.2024.138736}

\hypersetup{pdftitle={ATLAS document},pdfauthor={The ATLAS Collaboration}}
 
\begin{document}
 
\maketitle
 
\tableofcontents

\section{Introduction}

Leptoquarks (LQs) are hypothetical particles that carry both baryon
and lepton quantum numbers ($B\ne0, \ L\ne0$). They are either scalar (spin zero) or vector (spin
one) particles, carry colour and have a fractional electric charge.  LQs therefore couple
simultaneously to both quarks and leptons,
providing a connection between the lepton and quark sectors, which appear to have similar structures.
As such, LQs have already been discussed for many decades not only in unified theories~\cite{georgi_glashow_unification,pati_salam_colour, SU5_lq_constr}
but also in technicolour~\cite{dimop_suss,techni2,techni3} or composite models~\cite{composite_models_initial_1, composite_models_initial_2, SCHREMPP1985101}.
 
In the past decade, there has been renewed interest in LQs because of anomalies in various measurements of $B$-meson
decays.
While a recent reanalysis of data from the LHCb Collaboration has yielded
results in agreement with lepton-flavour universality~\cite{LHCb:2022qnv, LHCb:2022zom},
other anomalies such as in the $b \rightarrow s \mu \mu$ angular distribution or
in charged-current $b \rightarrow c \ell \nu$ transitions remain~\cite{babar,belle,lhcb1,Aaij:2015oid,Hirose:2016wfn,Belle:2016ure,Wehle:2016yoi,LHCb:2017smo,Aaij:2017tyk}.

This Letter combines searches from the ATLAS experiment for pair
production of LQs that decay into a charged or neutral lepton
and either a top or bottom quark. All analyses use the full \RunTwo
data set with an integrated luminosity of 139~fb$^{-1}$. This dataset was recorded
between 2015 and 2018 with the ATLAS detector~\cite{PERF-2007-01},
when the Large Hadron Collider (LHC) collided protons at $\sqrt{s} =
13$~\TeV.
 
\begin{figure}
\centering
\begin{subfigure}[b]{0.5\textwidth}
\centering
\includegraphics[width=0.9\linewidth]{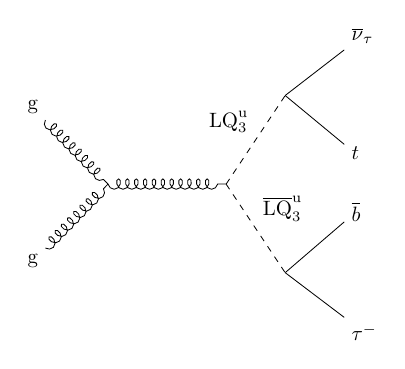}
\caption{}
\label{fig:ggF}
\end{subfigure}\begin{subfigure}[b]{0.5\textwidth}
\centering
\includegraphics[width=0.9\linewidth]{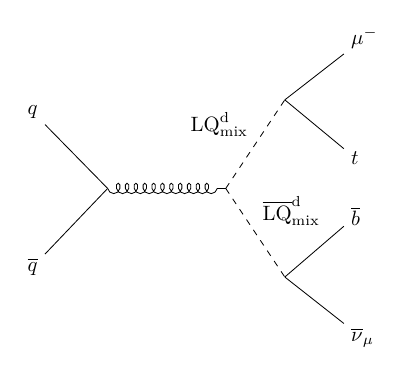}
\caption{}
\label{fig:qqA}
\end{subfigure}
\caption{Pair production of \subref{fig:ggF} up-type LQs of the third generation via gluon--gluon fusion with decays into a top quark and a neutrino or into a bottom quark and a $\tau$-lepton, and of \subref{fig:qqA} down-type mixed LQs via quark--antiquark annihilation with decays into a top quark and a muon or into a bottom quark and a neutrino.}
\label{fig:feynman}
\end{figure}

LQs decay into a quark and lepton of the same or of different
generations~\cite{Buchmuller:1986zs}.  Third-generation scalar LQs
have a charge of either $\pm 1/3$ ($\mathrm{LQ}_3^{\mathrm{d}}$) with
decays into $\bar{b}\bar{\nu}_\tau$ or $\bar{t}\tau^+$ (and charge
conjugate decays) or $\pm 2/3$ ($\mathrm{LQ}_3^{\mathrm{u}}$) with
decays into $b\tau^+$ or $t\bar{\nu}_\tau$ (and charge conjugate
decays). Mixed-generation scalar LQs can decay into a first- or second-generation
lepton and a third-generation quark. Similarly to $\mathrm{LQ}_3$, they can have charge
$\pm 1/3$  ($\mathrm{LQ}_{\mathrm{mix}}^{\mathrm{d}}$) or
$\pm 2/3$  ($\mathrm{LQ}_{\mathrm{mix}}^{\mathrm{u}}$).
Pairs of LQs would be produced at the LHC mainly via gluon--gluon fusion or quark--antiquark
annihilation. These mechanisms are shown in \Cref{fig:feynman}, together with some example decays.
The pair-production cross-section for scalar LQs depends only on the LQ mass ($m_{\LQ})$.

This Letter
considers nine independent searches that between them are sensitive to
all decays into a bottom or top quark and any one of the charged or neutral
leptons of all three generations. Sets of these analyses are combined to
produce two-dimensional exclusion limits for both third-generation and mixed scalar
LQs, as a function of $m_{\LQ}$ and branching ratio
($\mathcal{B}$) to the charged lepton in order to obtain
limits that are more stringent than those from the individual analyses.
A limited set of searches is also interpreted in vector LQ
models~\cite{Baker:2019sli}. The pair-production cross-section for vector LQs depends
not only on $m_{\LQ}$ but also on the coupling to gluons. An
additional coupling of Yang--Mills type to gluons is either present
or absent (\enquote{minimal coupling})~\cite{Bostonmodel:vLQ:1706.05033}.

It should be noted that because the ATLAS analyses included here do not have
any requirements on the LQ charge, the results for decays to a charged lepton are also
valid for LQs with charge $\pm 4/3$, e.g.\ with decays into
$\bar{b}\tau^+$,  or charge $\pm 5/3$, e.g.\ with decays into $t\tau^+$.


\section{Signal models}
 
Simulated events with pair-produced scalar up-type and down-type LQs were generated at next-to-leading order (NLO) in quantum chromodynamics (QCD) with \MGNLO[2.6.0]~\cite{Alwall:2014hca}. The method described in Ref.~\cite{Mandal:2015lca} is used, in which fixed-order NLO QCD calculations~\cite{Kramer:1997hh,Kramer:2004df} are interfaced to \PYTHIA[8.230]~\cite{Sjostrand:2014zea} for the parton shower (PS) and hadronisation.
Parton luminosities were provided by the five-flavour scheme \NNPDF[3.0nlo]~\cite{Ball:2014uwa_custom} parton distribution functions  with $\alphas=0.118$, and the underlying event (UE) was modelled with the A14 set of tuned parameters~\cite{ATL-PHYS-PUB-2014-021}.
\MADSPIN~\cite{Artoisenet:2012st} was used for the decay of the \LQ.
The branching ratio of LQ decays to a charged lepton is controlled by the model parameter $\beta$, so for
$\beta = 0$ there are only  decays into a neutrino and a quark, and  for $\beta = 1$ there are only decays into a charged lepton and a quark.
However, for equal couplings to charged and neutral leptons, i.e.\ $\beta = 0.5$, $\mathcal{B}$ differs from $0.5$ due to the sizeable top-quark mass~\cite{EXOT-2017-30}.
Signal samples were mostly produced for a model parameter of $\beta=0.5$, and target branching fractions $\mathcal{B}$ were obtained by reweighting the samples using generator-level information, as described in Ref~\cite{EXOT-2017-30}.
The coupling strength parameter $\lambda$ was set to $0.3$, resulting in a \LQ width of about \SI{0.2}{\percent} of its mass~\cite{Buchmuller:1986zs,Belyaev:2005ew}.

Signal cross-sections for scalar LQs were obtained from a calculation of the pair production of scalar coloured particles, e.g.\ the supersymmetric partner of the top quark. Such particles have the same production modes (except lepton t-channel contributions) and their pair-production cross-section depends only on their mass.
These cross-sections are computed at approximate next-to-next-to-leading order (NNLO) in QCD with resummation of next-to-next-to-leading logarithmic (NNLL) soft gluon terms~\cite{Beenakker:2016lwe,Beenakker:1997ut,Beenakker:2010nq,Beenakker:2016gmf}.
Lepton $t$-channel contributions are neglected and may lead to corrections at the percent level~\cite{Borschensky:2020hot}.

In addition to scalar LQs, two models of vector LQs are investigated.
Simulated events with pair-produced \UOne and \UOneTilde LQs were generated at leading order (LO) in QCD with \MGNLO, handling PDFs and the UE in the same way as for scalar LQs.
For \UOne LQs, a simplified model~\cite{Baker:2019sli} is used, neglecting further degrees of freedom relevant for the ultraviolet (UV) completion of the model.
The samples were produced with a coupling strength parameter of $g_U = 3.0$, leading to a signal width of approximately \SI{11}{\percent}~\cite{EXOT-2019-12}.
The large value of $g_U$ is motivated by a suppression of the production cross-section for additional mediators in an UV-complete model, which might otherwise be in tension with existing LHC $Z^\prime$ limits~\cite{Bostonmodel:vLQ:1706.05033}.
The coupling to gluons is determined through the chosen value of the model parameter $\kappa$, where $\kappa = 0$ corresponds to the Yang--Mills coupling scenario (\UOneYM) and $\kappa = 1$ to the minimal coupling scenario (\UOneMC).
The coupling strength to left- and right-handed quarks and leptons of the various generations is determined by the model parameters $\beta_\textrm{L\,/\,R}$.
While right-handed couplings are assumed to vanish in this analysis, $\beta_\textrm{L}$ is set to unity, resulting in equal coupling strengths of \UOne LQs to top quarks and neutrinos or bottom quarks and charged leptons of a certain generation.
Conversely, \UOneTilde LQs~\cite{Bostonmodel:vLQ:1706.05033, Bostonmodel:vLQ:1810.10017} carry an electric charge of $\pm 5/3$, and thus only decays into a top quark and $\tau$-lepton are allowed.
Couplings of \UOneTilde to the Standard Model gluon are governed by the model parameter $\kappa_t$.
Additional couplings of \UOneTilde to the heavy gluon $g'$, required for the UV completion of the model, depend on the model parameter $\kappa_s$.
The choice of $\kappa_t = \kappa_s = 0$ corresponds to the Yang--Mills scenario (\UOneTildeYM), whereas $\kappa_t = \kappa_s = -1$ gives the minimal coupling scenario (\UOneTildeMC).
Since production via an intermediate $g'$ contributes only in quark-initiated processes, which are subdominant compared to gluon--gluon fusion, the pair-production cross-section of \UOneTilde LQs depends only very slightly on the heavy-gluon mass, which is assumed to vanish in this analysis.
No higher-order cross-section computations are available for the \UOne and \UOneTilde models.
Therefore, the cross-sections computed at LO by \MGNLO are used in the analysis.

Additional details of the Monte Carlo simulation and the estimation of background processes in the individual analyses can be found in the respective publications.


\section{Experimental signatures}
In the following, each of the analyses contributing to the combined
results is described briefly, with further details available in the
references.  The final state in all analyses is reconstructed from
electrons~\cite{EGAM-2018-01}, muons~\cite{MUON-2018-03}, hadronically decaying $\tau$-leptons
(\tauhad)~\cite{PERF-2013-06}, jets and missing transverse momentum
(with magnitude \met)~\cite{PERF-2016-07}. Different quality criteria, isolation requirements and
transverse momentum (\pT) thresholds are applied in each analysis
because of the different backgrounds present. No
attempt is made to distinguish between a leptonically decaying $\tau$-lepton
and a prompt electron or muon ($\ell$). Jets are reconstructed from energy deposits in the calorimeter
system~\cite{PERF-2014-07} or from particle-flow
objects~\cite{PERF-2015-09,JETM-2018-05}, using the anti-$k_t$
algorithm~\cite{Cacciari:2008gp, Fastjet} with a radius parameter of 0.4 by default.
Jets can be categorised as
$b$-tagged (i.e.\ as \enquote{$b$-jets}) if they satisfy the requirements of a multivariate
algorithm~\cite{FTAG-2018-01,PERF-2016-05,ATL-PHYS-PUB-2017-013}. In
order to better reconstruct high-\pT top quarks, jets with a
larger radius-parameter value are used in some analyses, with the top-quark decay products
reconstructed from jet substructure information.

In most of the analyses, one or both LQs have a neutrino in the final
state, so they cannot be fully reconstructed.  Various methods are
employed to separate background from signal,
such as the usage of event-level variables designed to identify one or more missing particles, or
multivariate techniques that combine several final-state variables in an
optimised way.
In addition to signal-enriched regions, also called signal regions (SRs), control
regions (CRs) are defined for the purpose of constraining major background processes.
CRs and SRs are designed to be mutually exclusive.

\subsection*{Third-generation quarks and leptons}
 
\paragraph{$t\nu b\tau$~\cite{SUSY-2019-18}:}
The main selection requirements for this search are one \tauhad, two $b$-jets, and large \met, allowing searches in the
$\LQu{3}\rightarrow t\nu$
or  $ b\tau$ and
$\LQd{3}\rightarrow t\tau$ or
$b\nu$
channels.
The \pT of the \tauhad is used as the discriminant.
This channel is also interpreted in the $U_1^\mathrm{YM}$ and $U_1^\mathrm{MC}$ models.

\paragraph{$b\tau b\tau$~\cite{EXOT-2021-15_custom}:}
This search targets the  $\LQu{3}\rightarrow b\tau$  channel by
selecting events with either two or one \tauhad and one light charged
lepton (arising from a leptonic $\tau$ decay), together
with one or more $b$-jets.  The discriminant is the output of a parameterised
neural network~\cite{Baldi:2016fzo}, which uses as input various kinematic
variables including the partially reconstructed LQ masses.
This channel is also interpreted in the $U_1^\mathrm{YM}$ and $U_1^\mathrm{MC}$ models.

\paragraph{$t\tau t\tau$~\cite{EXOT-2019-15}:}
The search targets the $\LQd{3} \rightarrow t\tau$ channel.  Events are selected by
requiring between one and three light charged leptons,
at least one \tauhad and two or more jets, one of which is $b$-tagged.
The phase space of interest is divided into seven mutually exclusive SRs,
based on the light charged-lepton and
\tauhad multiplicities.  In each SR, the
effective mass, \meff, defined as the scalar sum of \met and the
transverse momenta of all reconstructed leptons and jets is used as the discriminant.
This channel is also reinterpreted in the
$\tilde{U}_1^\mathrm{YM}$ and $\tilde{U}_1^\mathrm{MC}$ models.

\subsection*{Third-generation quarks and first- or second-generation leptons}
 
\paragraph{$t\nu b\ell$~\cite{EXOT-2019-12}:}
The strategy of this search is to select events with one light charged
lepton, large \met and
four or more jets, of which at least one must be $b$-tagged.
It is therefore sensitive to
the  $\LQumix\rightarrow t\nu$, $be$
or $b\mu$ and
$\LQdmix \rightarrow  b\nu $, $te$
or $t\mu $ channels.  The
discriminant is the output of a neural network with
kinematic variables as input.
This channel is also interpreted in the
$U_1^\mathrm{YM}$ and $U_1^\mathrm{MC}$ models.

\paragraph{$b\ell b\ell$~\cite{EXOT-2019-13}:}
This analysis targets either the
$\LQumix \rightarrow be$ or
$b\mu$ channels.
Events are required to contain exactly two electrons or two muons with
opposite electric charges and at least two jets.  Events are then
categorised according to the multiplicity of the $b$-tagged jets.  LQ
candidates are identified by choosing the combination of leptons and
jets resulting in the smallest difference between the two lepton--jet
invariant masses. The average mass of the two LQ candidates
serves as the  discriminant.
 
\paragraph{$t\ell t\ell$ ($2\ell$)~\cite{EXOT-2019-19}:}
The search targets the $\LQdmix \rightarrow te$ or $t\mu$ channels.  The
analysis is optimised for signal events with both top quarks decaying
hadronically by requiring exactly two opposite-sign same-flavour
leptons and two $R=1.0$ jets containing the top decay products.
A boosted decision tree classifier (BDT) is trained to separate signal events from background events.  In
addition to variables calculated from the \pT of leptons and jets as
well as \met, observables obtained using recursive jigsaw
reconstruction techniques~\cite{Jackson:2017gcy} serve as input to the BDT.  The BDT output
score is used as the discriminant in the SRs.

\paragraph{$t\ell t\ell$ ($\geq 3\ell$)~\cite{ATLAS-CONF-2022-052}:}
Complementing the $t\ell t\ell$ ($2\ell$) analysis, this search
targets the same signal hypotheses in final states with at least three
light charged leptons. In addition, events are required to have at least two reconstructed jets,
at least one of which must be $b$-tagged. The SR is split on the
basis of lepton multiplicity, and \meff is used as the
discriminant.

\subsection*{Third-generation quarks and neutrinos}
 
\paragraph{$t\nu t\nu$~\cite{SUSY-2018-12}:}
The search for pair-produced scalar partners of the top quark in final
states consisting of two hadronically decaying top quarks and large
\met is also interpreted as a search for pair-produced LQs, primarily
targeting the $\LQu{3} \rightarrow t\nu$ channel. The main selection requirements in
the SR are four jets, two of which are $b$-tagged, no
charged leptons, and large \met. Top candidates are reconstructed using jets
with $R=1.2$.
Two such top candidates are required, and the SR is subdivided according to both the mass
of the top candidate with lower transverse momentum and the kinematic variable
$m_{\textrm{T2},\chi^2}$~\cite{Barr:2003rg}, which is a transverse mass suited to the
case where there are two missing particles.
For this publication, this channel is also reinterpreted in the
\LQumix, $U_1^\mathrm{YM}$ and $U_1^\mathrm{MC}$ models.

\paragraph{$b\nu b\nu$~\cite{SUSY-2018-34}:}
The search for scalar partners of the bottom quark in events with no light
charged leptons, two $b$-tagged jets and large \met also targets pair-produced LQs in the
$\LQd{3} \rightarrow b\nu$ channel. However, since there is neither a requirement on additional jets
nor explicit $\tau$ reconstruction the search also has sensitivity to
the $\mathrm{LQ}_{\mathrm{3}}^{\mathrm{d}} \rightarrow t\tau$ channel. A cut-based approach is used for high values
of a specific mass reconstructed in the presence of neutrinos, $m_\textrm{CT}$~\cite{Tovey:2008ui}.
The SR is subdivided into regions depending on $m_\textrm{CT}$ and \meff. At lower values of $m_\textrm{CT}$ a BDT
is used as the discriminant, with input variables based on the kinematics of the final-state jets and \met.
For this publication, this channel is also reinterpreted in the \LQdmix models.

\bigskip
A summary of the channels considered in this Letter, together with
which interpretations are made and the main selection requirements, is
given in Table~\ref{tab:analyses_overview}.
 
\begin{table}[ht] \scriptsize \centering
\caption{Overview of the individual analyses, together with the models used to interpret them and the main  object selections
in the signal region: number of  electrons or muons ($N_{\ell}$), number of hadronically decaying $\tau$-leptons
($N_{\tau_{\mathrm{had}}}$) and number of $b$-jets  ($N_{b \mathrm{jets}}$). All analyses interpreted in the same model are combined except for  $U_1$ in the $t\nu b \ell$ final state, where no other channels are available for combination.}
\label{tab:analyses_overview}
\begin{tabular}{ccccccccccc} \toprule & &
\multicolumn{6}{c}{\textbf{Interpretation}} & & & \\
\multicolumn{2}{c}{\textbf{Search}} &
\multicolumn{4}{c}{\textbf{Scalar}} &
\multicolumn{2}{c}{\textbf{Vector}} &
\multicolumn{3}{c}{\textbf{Signal Region}} \\ Final State & Citation  & $\mathrm{LQ}_{3}^{\mathrm{u}}$ &
$\mathrm{LQ}_{\mathrm{3}}^{\mathrm{d}}$ &
$\mathrm{LQ}_{\mathrm{mix}}^{\mathrm{u}}$ &
$\mathrm{LQ}_{\mathrm{mix}}^{\mathrm{d}}$ &
$U_1^{\mathrm{YM}/\mathrm{MC}}$ &
$\tilde{U}_1^{\mathrm{YM}/\mathrm{MC}}$ & $N_{\ell}$ &
$N_{\tau_{\mathrm{had}}}$ & $N_{b \mathrm{jets}}$ \\
\midrule
 
$ t\nu b\tau$ & \cite{SUSY-2019-18} &  \checkmark & \checkmark & -- & -- & \checkmark & -- & 0 & 1 & $\ge2$ \\
$b\tau b\tau$ & \cite{EXOT-2021-15_custom} &  \checkmark & -- & -- & -- & {\checkmark} & -- & $\{0,1\}$ & $\{1,2\}$ & $\{1,2\}$ \\
$t\tau t\tau$ & \cite{EXOT-2019-15} &  -- & \checkmark & -- & -- & -- & {\checkmark} & $\{1,2,3\}$ & $\ge1$ & $\ge1$ \\ \hline
$t\nu b \ell$ & \cite{EXOT-2019-12} &  -- & -- & \checkmark & \checkmark & \checkmark & -- & 1 & -- & $\ge1$ \\
$b\ell b\ell$ &  \cite{EXOT-2019-13} &  -- & -- & \checkmark & -- & -- & -- & 2 & -- & $\{0,1,2\}$ \\
$t\ell t\ell$ ($2 \ell$) & \cite{EXOT-2019-19} &  -- & -- & -- & \checkmark & -- & -- & 2 & -- & -- \\
$t\ell t\ell$ ($\ge 3 \ell$) & \cite{ATLAS-CONF-2022-052} &  -- & -- & -- & \checkmark & -- & -- & $\{3,4\}$ & -- & $\ge2$ \\ \hline
$t\nu t \nu$ & \cite{SUSY-2018-12} &  \checkmark & -- & {\checkmark} & -- & {\checkmark} & -- & 0 & 0 & $\ge2$ \\
$b\nu b \nu$ & \cite{SUSY-2018-34} &  -- & \checkmark & -- & {\checkmark} & -- & -- & 0 & -- & $\ge2$ \\

\bottomrule
\end{tabular}
\end{table}


\section{Statistical interpretation}
\label{interpretation}
The statistical combination of the different searches follows the same formalism as the individual analyses.
Simultaneous binned profile-likelihood fits to CRs and SRs, steered by the \texttt{cabinetry} library~\cite{cabinetry_zenodo, cabinetry_chep}, are performed, and a modified frequentist method, the CL$_\text{s}$ method~\cite{Read:2002hq}
as implemented in \texttt{pyhf}~\cite{pyhf, pyhf_joss}, is used to set exclusion limits on various signal hypotheses.
Systematic uncertainties are accounted for by including them in the fits via nuisance parameters (NPs).
The binned likelihood function $\mathcal{L}(\mu, \theta)$ is constructed as the product of Poisson probability terms over all bins considered in the analysis and a number of Gaussian or log-normal priors for the NPs associated with systematic uncertainties.
It depends on the signal strength parameter, $\mu$, a multiplicative factor applied to the theoretical signal production cross-section, and on $\theta$, a set of NPs including a number of unconstrained multiplicative factors for the major background processes as well as those entering the Gaussian and log-normal priors, which adjust the expectations for signal and background according to the corresponding systematic uncertainties.
 
The test statistic used to determine the exclusion limits is based on the profile likelihood ratio
\begin{align*}
t_{\mu} = -2 \ln{\frac{\mathcal{L}(\mu,\hat{\hat{\theta}})}{\mathcal{L}(\hat{\mu},\hat{\theta})}},
\end{align*}
where $\hat{\mu}$ and $\hat{\theta}$ are the values of the parameters that maximise the likelihood function, and $\hat{\hat{\theta}}$ are the values of the NPs that maximise the likelihood function for a given value of $\mu$.
Upper limits on the signal production cross-section are derived for each of the signal scenarios considered in this analysis.
For a given signal scenario, values of the production cross-section (parameterised by $\mu$) yielding CL$_\text{s} < 0.05$, where CL$_\text{s}$ is computed using the asymptotic approximation~\cite{Cowan:2010js}, are excluded at $\geq \SI{95}{\percent}$ confidence level (CL).

The sensitivity of each individual analysis entering the combination is limited primarily by statistical uncertainties.
As such, systematic uncertainties do not have a major impact in the combined analysis, and therefore assumptions about their correlation across the different analyses do not affect the results significantly.
Detector-related uncertainties are assumed to be correlated across analyses, unless that is not valid because of different approaches to  event reconstruction or a different choice of correlation model in individual analyses.
Similarly, uncertainties in the modelling of the signal process are only assumed to be correlated when parameterised in the same way in individual analyses.
However, treating these uncertainties as fully uncorrelated yields a difference of at most 5\% on the cross-section limit.
Uncertainties in the modelling and normalisation of background processes are assumed to be uncorrelated between analyses, since the phase spaces of interest and background estimation techniques can differ significantly.
As is standard in ATLAS searches, theory uncertainties affecting the signal acceptance are considered when setting exclusion limits, but those affecting the production cross-section are not.

In each combination, the SRs and CRs defined for the individual searches
are generally statistically independent between the various analyses,
due to the
selection requirements on the multiplicity and flavour of the charged leptons.
Additional criteria are used in the one-lepton CRs in the (otherwise zero-lepton) $t\nu t\nu$ channel, such as
the requirement that the lepton \pT is below the threshold used in the  SRs and CRs of the other channels.
Residual overlaps caused by the use of different lepton identification criteria were found to be negligible in all cases except
between one of the six SRs of the $t\nu t\nu$ analysis and the SR of the $t\nu b\tau$ analysis.
Since this SR in the $t\nu t\nu$ analysis does not significantly
affect the sensitivity for LQ signal processes, it is removed from the
combination.  Other overlaps between SRs of different analyses do not
exceed \SI{7}{\percent} (\SI{2}{\percent}), as estimated with Monte
Carlo signal samples (data), and have negligible impact on the results
presented in this Letter.  Similarly, statistical overlaps between CRs
in one analysis and SRs or CRs in other analyses have negligible
impact on the results and are therefore ignored.


\section{Results}

No significant deviations from the Standard Model expectation are observed in any of the analysis channels considered in this Letter, and good agreement is also found in the combined background-only fits to data.
Upper limits on the cross-sections for LQ pair production are calculated at \SI{95}{\percent} CL using simultaneous signal-plus-background fits to the CRs and SRs, in which the background normalisations and possible signal contributions are determined.
Such limits on the cross-sections are evaluated for a wide range of values for the branching ratio of LQs into charged leptons.
Comparisons with theoretical cross-section predictions provide mass exclusion limits.

\begin{figure}
\centering
\begin{subfigure}[b]{0.5\textwidth}
\centering
\includegraphics[width=\textwidth]{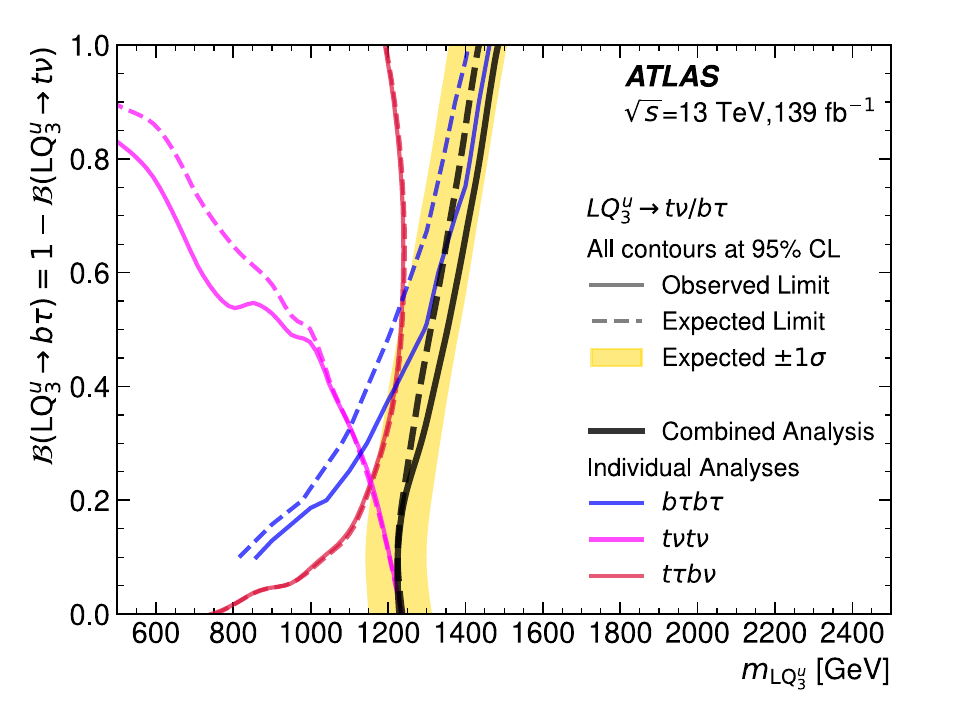}
\caption{}
\label{fig:results:contours:lq3u}
\end{subfigure}\begin{subfigure}[b]{0.5\textwidth}
\centering
\includegraphics[width=\textwidth]{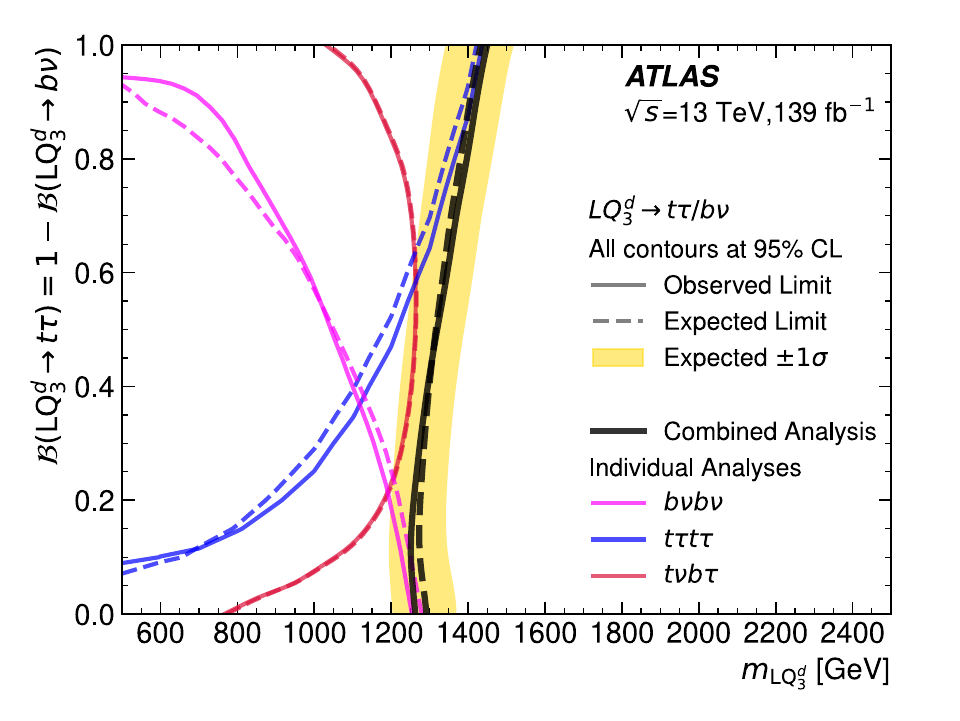}
\caption{}
\label{fig:results:contours:lq3d}
\end{subfigure}
\caption{Expected and observed \SI{95}{\percent} CL exclusion limits on the LQ mass as a function of the branching ratio to a charged lepton. Limits are presented for third-generation \subref{fig:results:contours:lq3u} up-type and \subref{fig:results:contours:lq3d} down-type LQs decaying into a third-generation quark and lepton. The full $\mathcal{B}$ parameter space is not evaluated for some analyses if they have limited sensitivity.}
\label{fig:results:contours:lq3}
\end{figure}

The exclusion limits for scalar third-generation up-type LQs obtained
from the combined analysis are shown in
\Cref{fig:results:contours:lq3u} on top of an overlay of the limits
from the three individual analyses.  As discussed in
Section~\ref{interpretation} the limits are dominated by statistical
uncertainties.  As an example, at $\mathcal{B} = 0.4$ and a LQ mass of 1.3~TeV the dominating
systematic uncertainties on the cross-section are due to background
modelling (7\%) and signal modelling (4\%).Compared to the individual
searches, the exclusion reach is improved by up to \SI{100}{\GeV} for
intermediate values of $\mathcal{B}$, where more than one individual
analysis provides sensitivity.  Similar observations are made in the
case of scalar third-generation down-type LQs, where the exclusion
limits improve by up to \SI{70}{\GeV} as shown in
\Cref{fig:results:contours:lq3d}.
 
\begin{figure}
\centering
\begin{subfigure}[b]{0.5\textwidth}
\centering
\includegraphics[width=\textwidth]{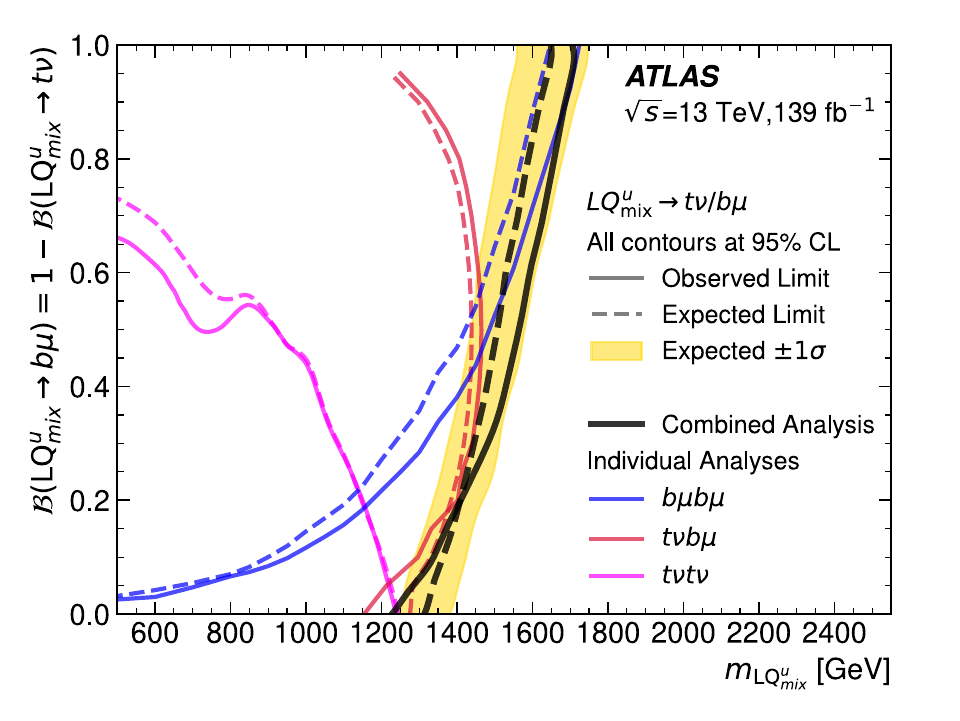}
\caption{}
\label{fig:results:contours:lqxumu}
\end{subfigure}\begin{subfigure}[b]{0.5\textwidth}
\centering
\includegraphics[width=\textwidth]{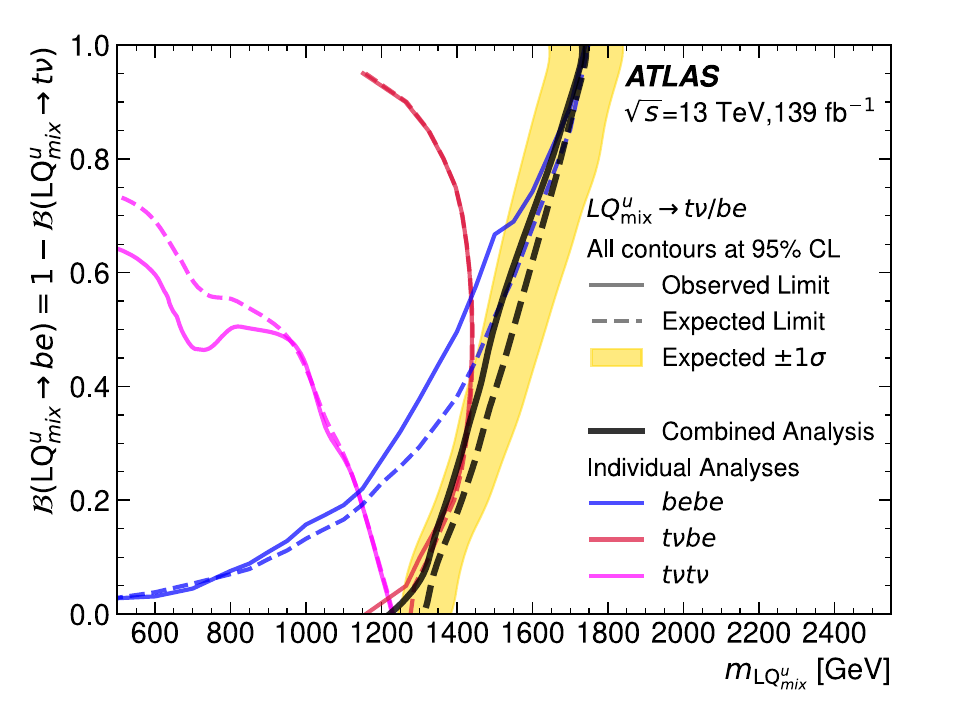}
\caption{}
\label{fig:results:contours:lqxuel}
\end{subfigure}
\caption{Expected and observed \SI{95}{\percent} CL exclusion limits on the LQ mass as a function of the branching ratio to a charged lepton. Limits are presented for up-type scalar LQs decaying into a third-generation quark and \subref{fig:results:contours:lqxumu} a muon or \subref{fig:results:contours:lqxuel}  an electron. The full $\mathcal{B}$ parameter space is not evaluated for some analyses if they have limited sensitivity.}
\label{fig:results:contours:lqxu}
\end{figure}

The combined exclusion limits also show improvement with respect to
the three individual analyses sensitive to scalar up-type LQs decaying
into a third-generation quark and a first- or second-generation lepton. This
is especially true for low and intermediate values of $\mathcal{B}$.
The corresponding exclusion contours are shown in \Cref{fig:results:contours:lqxu} for decays to a muon (a) or an electron (b), with improvements in the observed exclusion limits of up to \SI{80}{\GeV} and \SI{90}{\GeV}, respectively.
 
\begin{figure}
\centering
\begin{subfigure}[b]{0.5\textwidth}
\centering
\includegraphics[width=\textwidth]{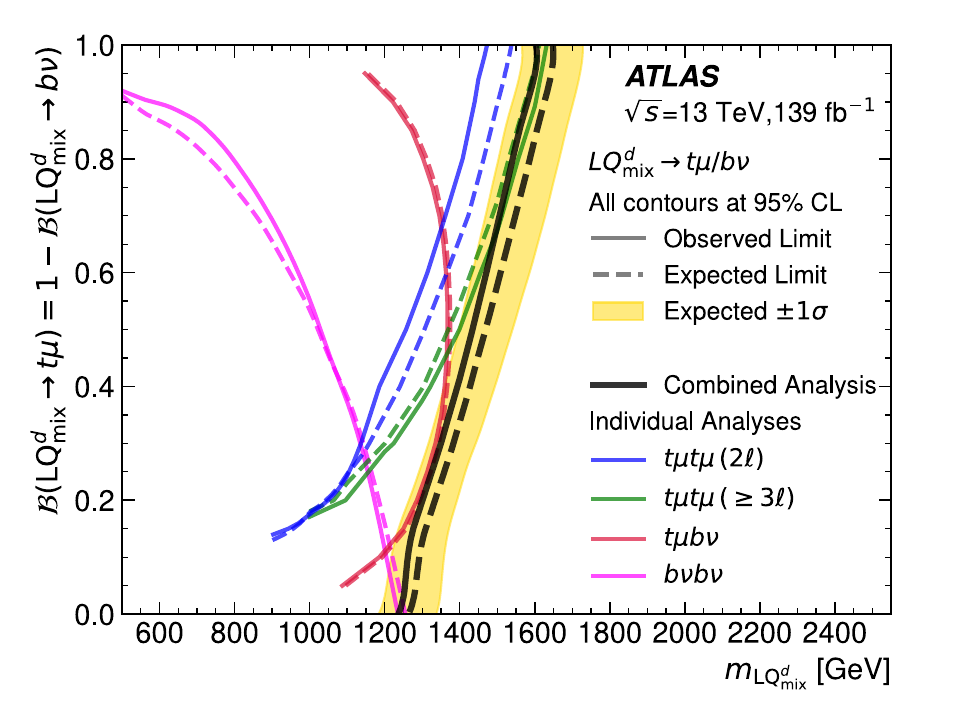}
\caption{}
\label{fig:results:contours:lqxdmu}
\end{subfigure}\begin{subfigure}[b]{0.5\textwidth}
\centering
\includegraphics[width=\textwidth]{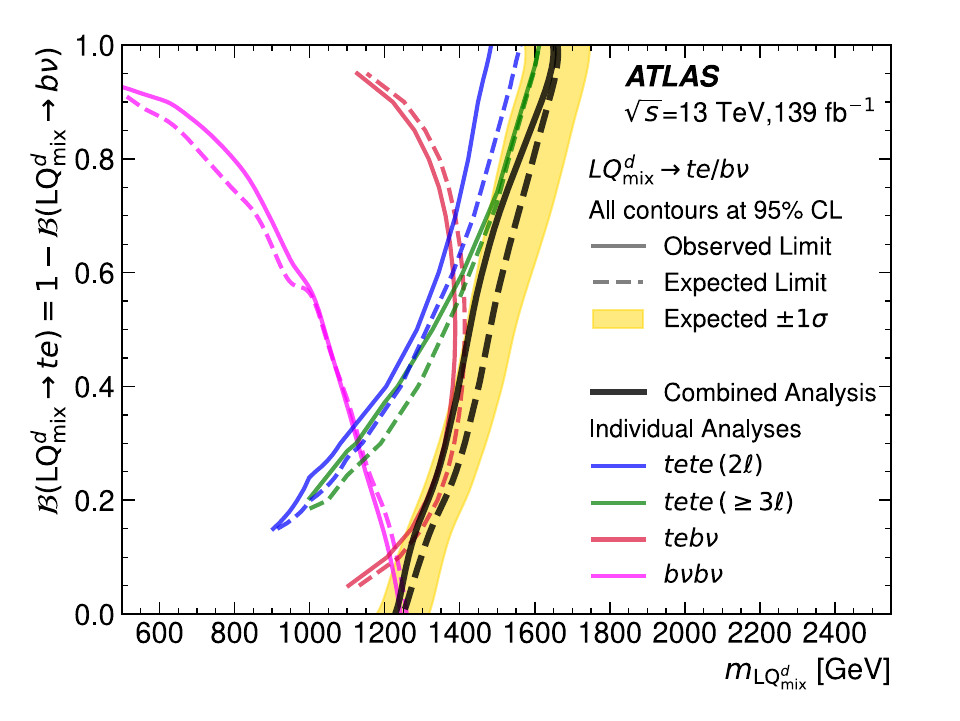}
\caption{}
\label{fig:results:contours:lqxdel}
\end{subfigure}
\caption{Expected and observed \SI{95}{\percent} CL exclusion limits on the LQ mass as a function of the branching ratio to a charged lepton. Limits are presented for down-type scalar LQs decaying into a third-generation quark and  \subref{fig:results:contours:lqxdmu} a muon or \subref{fig:results:contours:lqxdel} an electron. The full $\mathcal{B}$ parameter space is not evaluated for some analyses if they have limited sensitivity.}
\label{fig:results:contours:lqxd}
\end{figure}

In the case of scalar down-type LQs decaying into a third-generation quark and a first- or second-generation lepton, improvements in sensitivity are observed across the whole range of $\mathcal{B}$, since a total of four individual analysis channels are combined here, generally also resulting in higher exclusion limits when the LQ is assumed to decay exclusively into a top quark and a charged lepton.
Corresponding results are shown in \Cref{fig:results:contours:lqxd}, with the combination improving the exclusion limits by up to \SI{60}{\GeV} for decays to a muon and by up to \SI{80}{\GeV} for decays to an electron.
For LQs decaying into a top quark and a muon with $\mathcal{B} \approx 1$, the observed combined lower limit on the LQ mass is slightly weaker than the one observed in $t\mu t\mu\,(\geq 3\ell)$ since a worse-than-expected limit is observed in $t\mu t\mu\,(2\ell)$.
 
\begin{figure}
\centering
\begin{subfigure}[b]{0.5\textwidth}
\centering
\includegraphics[width=\textwidth]{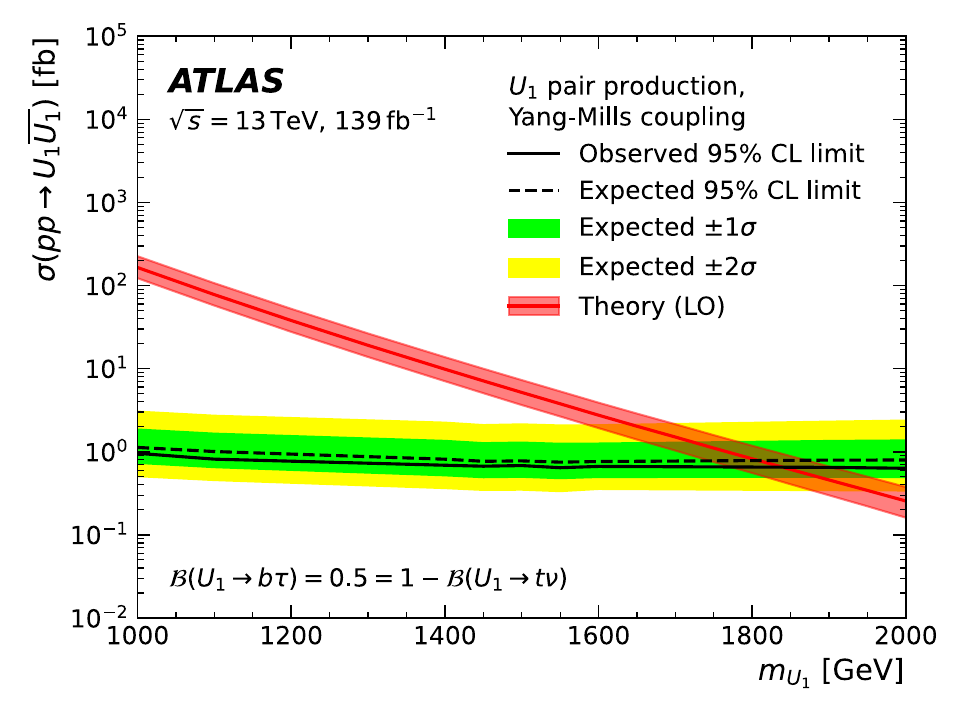}
\caption{}
\label{fig:results:limits1d:u1YM}
\end{subfigure}\begin{subfigure}[b]{0.5\textwidth}
\centering
\includegraphics[width=\textwidth]{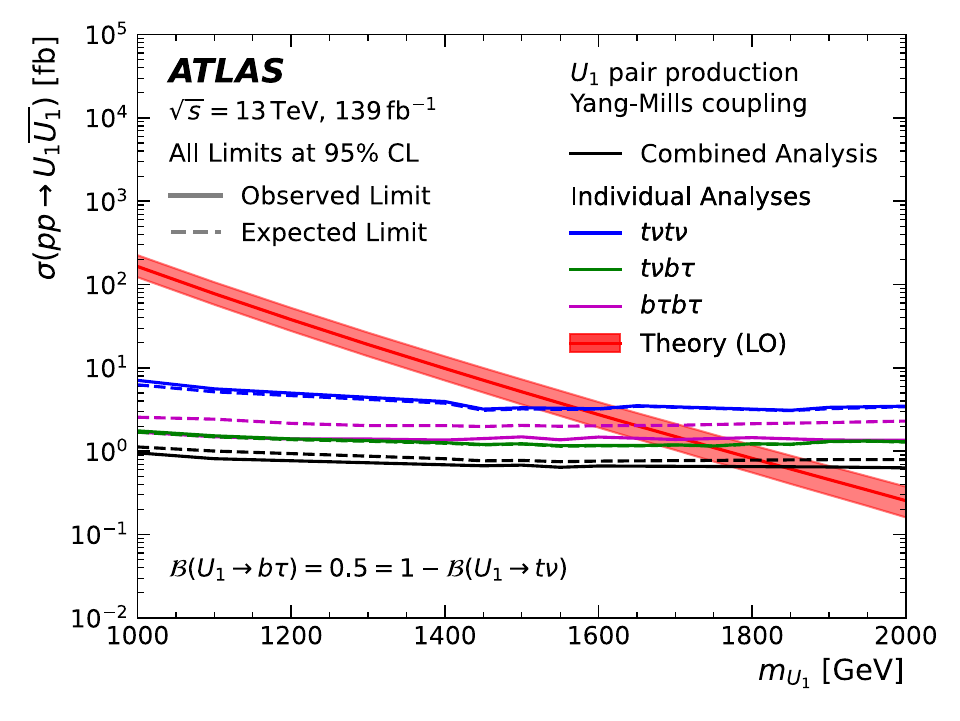}
\caption{}
\label{fig:results:limits1d:u1YMmulticurve}
\end{subfigure}
\caption{Expected and observed \SI{95}{\percent} CL exclusion limits on the \UOneYM production cross-section as a function of the LQ mass, with a branching ratio of 0.5 to a $\tau$-lepton. The limits obtained from the combined analysis, including uncertainty bands, are presented in \subref{fig:results:limits1d:u1YM}, whereas \subref{fig:results:limits1d:u1YMmulticurve} shows the limits obtained from the individual analyses for comparison. The theoretical prediction and its $\pm1\sigma$ uncertainty due to variations of the parton distribution functions, the strong coupling parameter and the renormalisation and factorisation scales are shown in red.}
\label{fig:results:limits1d:u1:ym}
\end{figure}

\begin{figure}
\centering
\begin{subfigure}[b]{0.5\textwidth}
\centering
\includegraphics[width=\textwidth]{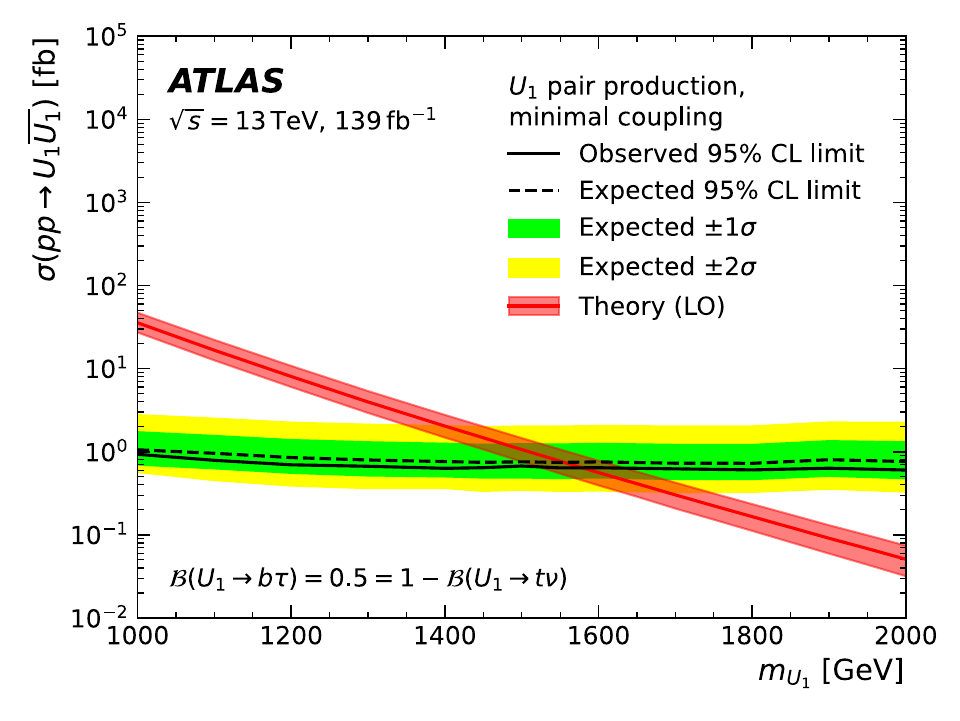}
\caption{}
\label{fig:results:limits1d:u1min}
\end{subfigure}\begin{subfigure}[b]{0.5\textwidth}
\centering
\includegraphics[width=\textwidth]{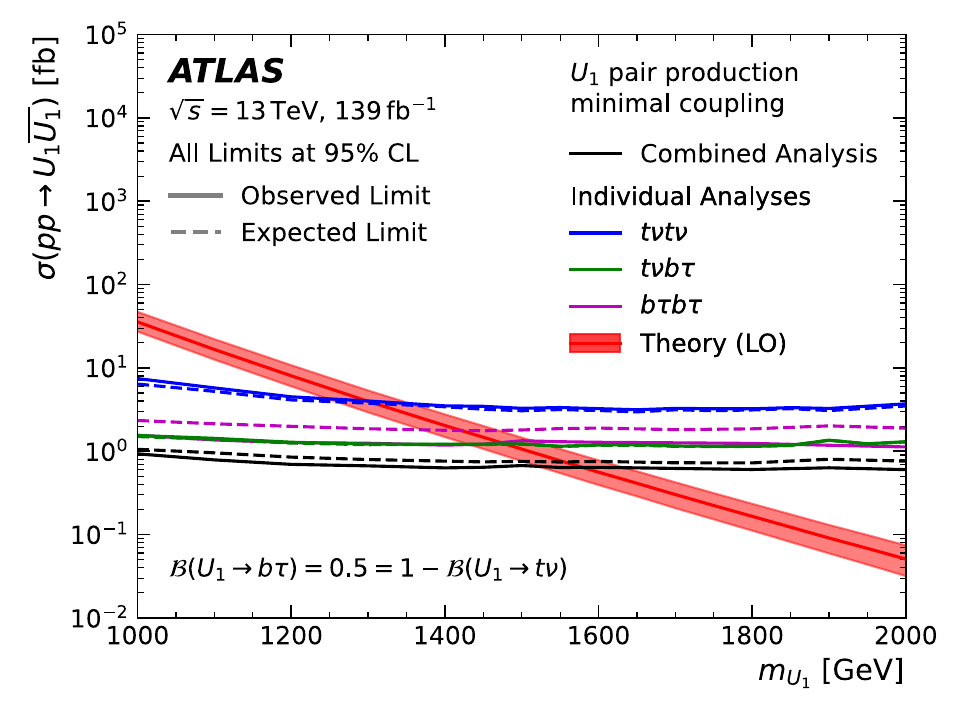}
\caption{}
\label{fig:results:limits1d:u1minmulticurve}
\end{subfigure}
\caption{Expected and observed \SI{95}{\percent} CL exclusion limits on the \UOneMC production cross-section as a function of the LQ mass with a branching ratio of 0.5 to a $\tau$-lepton. The limits obtained from the combined analysis, including uncertainty bands, are presented in \subref{fig:results:limits1d:u1min}, whereas \subref{fig:results:limits1d:u1minmulticurve} shows the limits obtained from the individual analyses for comparison. The theoretical prediction and its $\pm1\sigma$ uncertainty due to variations of the parton distribution functions, the strong coupling parameter and the renormalisation and factorisation scales are shown in red.}
\label{fig:results:limits1d:u1:min}
\end{figure}

In addition to mass exclusion limits for scalar LQs, limits on the production cross-section of \UOne vector LQs decaying into a third-generation quark and lepton are provided,
assuming $\mathcal{B}=0.5$.
The results of combining the three contributing analyses are shown in \Cref{fig:results:limits1d:u1:ym} and \Cref{fig:results:limits1d:u1:min}
for \UOneYM and \UOneMC, respectively.
The observed (expected) lower limits on the \UOne mass are \SI{1840}{\GeV} (\SI{1810}{\GeV}) for \UOneYM
and \SI{1580}{\GeV} (\SI{1560}{\GeV}) for \UOneMC.
For these signal hypotheses, the combined expected limits are primarily driven by the searches in the $t\nu b\tau$ final state as expected.
The combined observed limits are slightly higher than the expected ones, mainly due to the $b\tau b\tau$ channel showing similar behaviour, and are improved by \SI{70}{\GeV} (\UOneYM) and \SI{80}{\GeV} (\UOneMC) compared to the individual analyses.

\begin{figure}
\centering
\begin{subfigure}[b]{0.5\textwidth}
\centering
\includegraphics[width=\textwidth]{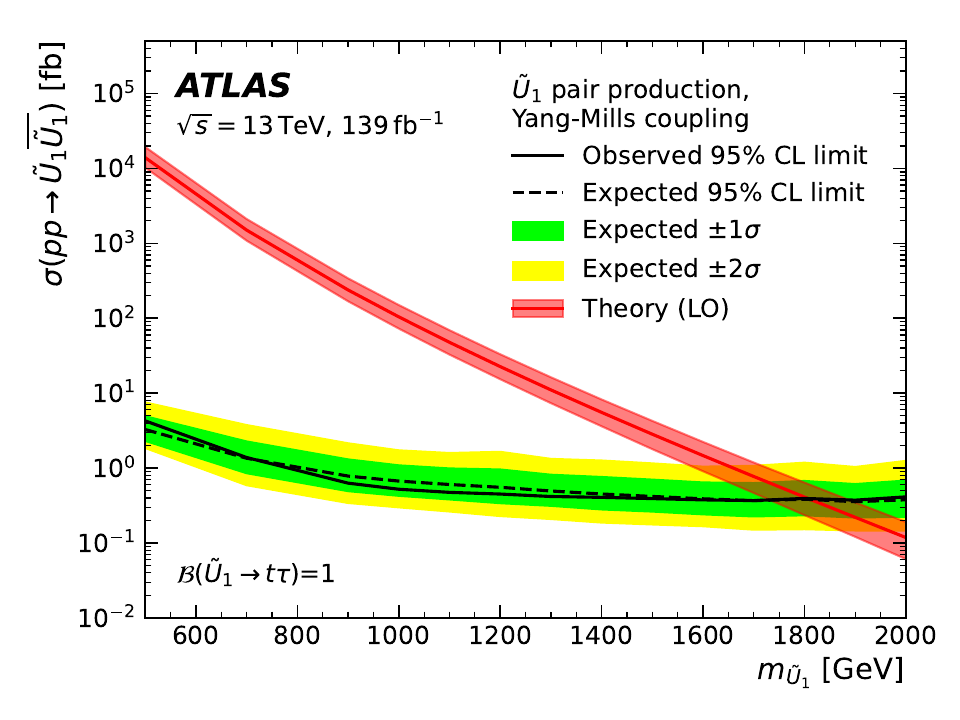}
\caption{}
\label{fig:results:limits1d:u1tildeYM}
\end{subfigure}\begin{subfigure}[b]{0.5\textwidth}
\centering
\includegraphics[width=\textwidth]{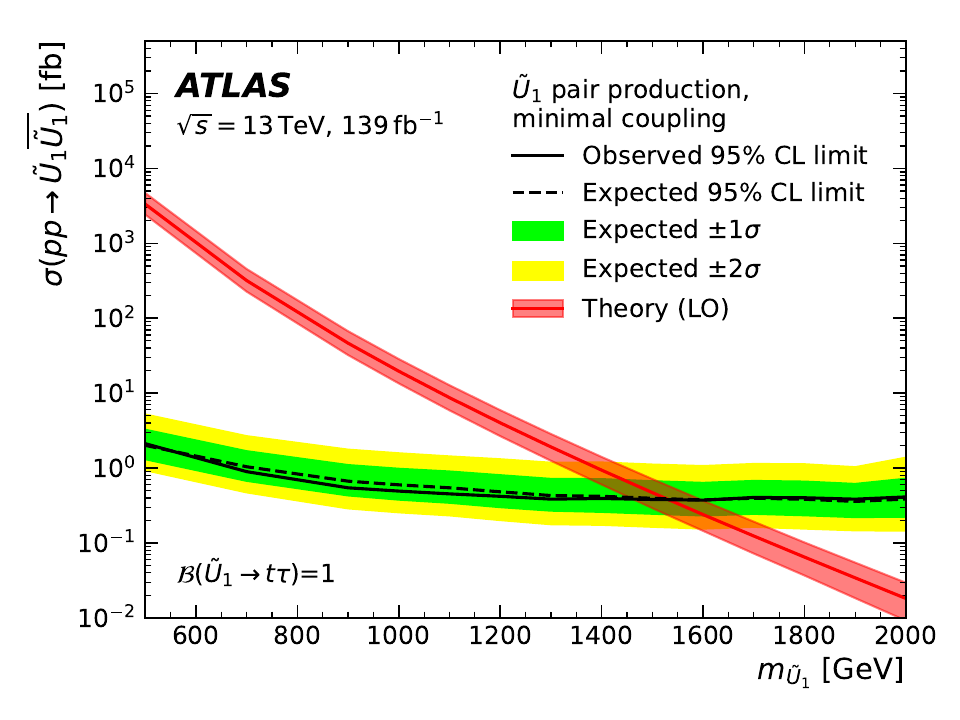}
\caption{}
\label{fig:results:limits1d:u1tildemin}
\end{subfigure}
\caption{Expected and observed \SI{95}{\percent} CL exclusion limits on the \UOneTilde production cross-section as a function of the LQ mass with a branching ratio of 1.0 to a $\tau$-lepton. Limits are presented for \subref{fig:results:limits1d:u1tildeYM} \UOneTildeYM and  \subref{fig:results:limits1d:u1tildemin} \UOneTildeMC. The theoretical prediction and its $\pm1\sigma$ uncertainty due to variations of the parton distribution functions, the strong coupling parameter and the renormalisation and factorisation scales are shown in red.}
\label{fig:results:limits1d:u1tilde}
\end{figure}

Finally, the search in \ttauttau final states is reinterpreted in the \UOneTilde vector LQ model.
Since \UOneTilde carries an electric charge of $\pm 5/3$, it decays exclusively into a top quark and a $\tau$-lepton.
The resulting upper limits on the production cross-section are shown in \Cref{fig:results:limits1d:u1tilde}, with observed (expected) lower limits on the \UOneTilde mass of \SI{1810}{\GeV} (\SI{1810}{\GeV}) and \SI{1540}{\GeV} (\SI{1530}{\GeV}) for \UOneTildeYM and \UOneTildeMC, respectively.
 
Table~\ref{tab:summary} summarises the lower limits on the LQ mass
obtained in this analysis at $\mathcal{B} = 0.0, 0.5, 1.0$,
including four additional limits~\cite{EXOT-2019-12} on \UOne
vector LQs that decay into a third-generation quark and a first- or second-generation
lepton, where no other channels are available for a
combination.  Since the analyses are not sensitive to the quark charge,
the results shown for scalar LQs with $\mathcal{B}=1$, i.e.\ the last
column and first six rows of Table~\ref{tab:summary}, are also valid
for scalar LQs with charge $\pm 4/3$ and $be$,
$b\mu$ or $b\tau$ decays or with charge $\pm 5/3$ and $te$, $t\mu$, or $t\tau$ decays.

\begin{table}
\centering
\caption{Expected and observed \SI{95}{\percent} CL lower limits on the LQ mass at branching ratios to a charged lepton of $\mathcal{B} = 0.0, 0.5, 1.0$ for the signal hypotheses considered in this analysis, including four additional limits~\cite{EXOT-2019-12}
on the \UOne
vector LQs decaying into a third-generation quark and a first- or second-generation lepton,
where no other channels are available for a combination.}
 
{\def\arraystretch{1.2}
\begin{tabular}{l|cc|cc|cc}
\toprule
& \multicolumn{2}{c|}{$\mathcal{B}=0.0$}                 & \multicolumn{2}{c|}{$\mathcal{B}=0.5$}                 & \multicolumn{2}{c}{$\mathcal{B}=1.0$}                 \\
& \multicolumn{2}{c|}{\SI{95}{\percent} CL Limit [\text{G\electronvolt}]} & \multicolumn{2}{c|}{\SI{95}{\percent} CL Limit [\text{G\electronvolt}]} & \multicolumn{2}{c}{\SI{95}{\percent} CL Limit [\text{G\electronvolt}]} \\
& Observed              & Expected                       & Observed              & Expected                       & Observed             & Expected                       \\
\midrule
$\LQu{3} \rightarrow t\nu/b\tau$ 	& 1240 & $1240_{-90}^{+70}$	                          & 1340	& $1300_{-80}^{+70}$	 		         & 1480 & $1440_{-80}^{+70}$	 	                          \\
$\LQd{3} \rightarrow t\tau/b\nu$ 	& 1260 & $1260_{-80}^{+80}$	                        & 1360  & $1340_{-70}^{+60}$	 			         & 1520 & $1470_{-70}^{+70}$	 	                          \\
$\LQumix \rightarrow t\nu/b\mu$ 	& 1230 & $1310_{-70}^{+70}$	                       & 1570    & $1510_{-70}^{+70}$	 			         & 1710 & $1650_{-90}^{+90}$	 	                          \\
$\LQumix \rightarrow t\nu/be$ 		& 1230 & $1310_{-70}^{+70}$				         & 1510 & $1550_{-80}^{+80}$	 	                       & 1730   & $1740_{-100}^{+90}$	                            \\
$\LQdmix \rightarrow t\mu/b\nu$ 	& 1240 & $1260_{-80}^{+70}$				         & 1430 & $1470_{-70}^{+70}$	 	                        & 1600   & $1650_{-80}^{+80}$	                           \\
$\LQdmix \rightarrow te/b\nu$ 		& 1230 & $1250_{-70}^{+70}$				         & 1450 & $1500_{-70}^{+70}$	 	                         & 1650 & $1660_{-90}^{+90}$	                            \\
$\UOneYM \rightarrow t\nu/b\tau$        & -                     & -                              & 1840 & $1810_{-90}^{+80}$                              & -                    & -                              \\
$\UOneMC \rightarrow t\nu/b\tau$        & -                     & -                              & 1580 & $1560_{-70}^{+70}$                               & -                    & -                              \\
$\UOneYM \rightarrow t\nu/b\mu$	        & -     		& -                              & 1980 & $1930_{-60}^{+50}$                               & -                    & -                              \\
$\UOneMC \rightarrow t\nu/b\mu$		& - 			& -                              & 1710 & $1660_{-50}^{+50}$                               & -                    & -                              \\
$\UOneYM \rightarrow t\nu/be$    	& -	                & -                              & 1900 & $1930_{-70}^{+50}$                               & -                    & -                              \\
$\UOneMC \rightarrow t\nu/be$		& -     		& -                              & 1620 & $1650_{-60}^{+50}$                               & -                    & -                              \\
$\UOneTildeYM \rightarrow t\tau$   	& -                     & -                              &       -          	 &    -                        	& 1810   & $1810_{-70}^{+80}$                             \\
$\UOneTildeMC \rightarrow t\tau$   	& -                     & -                              &       -          	 &    -                        	 & 1540  & $1530_{-60}^{+90}$                             \\
\bottomrule
\end{tabular}
}
\label{tab:summary}
\end{table}
A comparison to results from the CMS collaboration is not directly
possible for results on vector LQs~\cite{CMS-EXO-19-015,CMS-EXO-19-016,CMS-B2G-16-027} due to the different
models used by ATLAS and CMS.
For the pair production of scalar LQs the individual limits are
already better than the CMS
limits~\cite{CMS-EXO-19-016,CMS-EXO-21-002,CMS-EXO-19-015,CMS-SUS-19-005,CMS-B2G-16-027,CMS-SUS-19-005}\footnote{After
submission of this paper a new result from the CMS Collaboration has
been reported~\cite{CMS:2024bnj}, yielding results on the search for
scalar and vector LQs in the $b\mu b\mu$ final state.}
at $\mathcal{B} = 0.0, 0.5, 1.0$ for all decay channels.
The limits shown in \Cref{fig:results:contours:lq3,fig:results:contours:lqxu,fig:results:contours:lqxd} therefore
constitute the best limits to date for any value of $\mathcal{B}$.


\FloatBarrier
 
\section{Conclusion}

A statistical combination of searches for pair-produced leptoquarks that decay into a
third-generation quark ($t$ or $b$) and any charged or neutral lepton
is presented. The analysis exploits the full data sample recorded with
the ATLAS detector in \RunTwo of the LHC, corresponding to
139\,fb$^{-1}$ of proton--proton collisions at $\sqrt{s} = 13$\,\TeV.
No significant deviation from the Standard Model expectation is
observed. At 95\% CL the lower limits on the masses of scalar LQs are found
to range from 1230 to 1730\,GeV, depending on branching ratio,
leptoquark charge and lepton flavour. The combined analyses extend the mass exclusion range by up to 100 GeV compared to the previous best individual analyses.
For any combination of these parameters the ATLAS limits constitute the best limits to date.
 
Limits are also placed on some benchmark models of vector LQs.
In the $U_1^{\mathrm{MC}}$ model, where
the coupling to gluons is minimal, LQs are excluded in the mass range below
1580 to 1710\,GeV for $\mathcal{B}=0.5$, depending on LQ flavour. For the
Yang--Mills model $U_1^{\mathrm{YM}}$, LQs are excluded in the mass
range below 1840 to 1980\,GeV for $\mathcal{B}=0.5$, depending on LQ
flavour. The combined analysis extends
the mass exclusion range by up to \SI{80}{\GeV} compared to the best individual analyses.
Vector LQs with charge $\pm 5/3$
and $t\tau$ decays are excluded up to 1540\,GeV for
$\tilde{U}_1^{\mathrm{MC}}$ and up to 1810\,GeV for $\tilde{U}_1^{\mathrm{YM}}$.


\section*{Acknowledgements}


We thank CERN for the very successful operation of the LHC and its injectors, as well as the support staff at
CERN and at our institutions worldwide without whom ATLAS could not be operated efficiently.
 
The crucial computing support from all WLCG partners is acknowledged gratefully, in particular from CERN, the ATLAS Tier-1 facilities at TRIUMF/SFU (Canada), NDGF (Denmark, Norway, Sweden), CC-IN2P3 (France), KIT/GridKA (Germany), INFN-CNAF (Italy), NL-T1 (Netherlands), PIC (Spain), RAL (UK) and BNL (USA), the Tier-2 facilities worldwide and large non-WLCG resource providers. Major contributors of computing resources are listed in Ref.~\cite{ATL-SOFT-PUB-2023-001}.
 
We gratefully acknowledge the support of ANPCyT, Argentina; YerPhI, Armenia; ARC, Australia; BMWFW and FWF, Austria; ANAS, Azerbaijan; CNPq and FAPESP, Brazil; NSERC, NRC and CFI, Canada; CERN; ANID, Chile; CAS, MOST and NSFC, China; Minciencias, Colombia; MEYS CR, Czech Republic; DNRF and DNSRC, Denmark; IN2P3-CNRS and CEA-DRF/IRFU, France; SRNSFG, Georgia; BMBF, HGF and MPG, Germany; GSRI, Greece; RGC and Hong Kong SAR, China; ISF and Benoziyo Center, Israel; INFN, Italy; MEXT and JSPS, Japan; CNRST, Morocco; NWO, Netherlands; RCN, Norway; MEiN, Poland; FCT, Portugal; MNE/IFA, Romania; MESTD, Serbia; MSSR, Slovakia; ARRS and MIZ\v{S}, Slovenia; DSI/NRF, South Africa; MICINN, Spain; SRC and Wallenberg Foundation, Sweden; SERI, SNSF and Cantons of Bern and Geneva, Switzerland; MOST, Taipei; TENMAK, T\"urkiye; STFC, United Kingdom; DOE and NSF, United States of America.
 
Individual groups and members have received support from BCKDF, CANARIE, CRC and DRAC, Canada; CERN-CZ, PRIMUS 21/SCI/017 and UNCE SCI/013, Czech Republic; COST, ERC, ERDF, Horizon 2020, ICSC-NextGenerationEU and Marie Sk{\l}odowska-Curie Actions, European Union; Investissements d'Avenir Labex, Investissements d'Avenir Idex and ANR, France; DFG and AvH Foundation, Germany; Herakleitos, Thales and Aristeia programmes co-financed by EU-ESF and the Greek NSRF, Greece; BSF-NSF and MINERVA, Israel; Norwegian Financial Mechanism 2014-2021, Norway; NCN and NAWA, Poland; La Caixa Banking Foundation, CERCA Programme Generalitat de Catalunya and PROMETEO and GenT Programmes Generalitat Valenciana, Spain; G\"{o}ran Gustafssons Stiftelse, Sweden; The Royal Society and Leverhulme Trust, United Kingdom.
 
In addition, individual members wish to acknowledge support from CERN: European Organization for Nuclear Research (CERN PJAS); Chile: Agencia Nacional de Investigaci\'on y Desarrollo (FONDECYT 1190886, FONDECYT 1210400, FONDECYT 1230812, FONDECYT 1230987); China: National Natural Science Foundation of China (NSFC - 12175119, NSFC 12275265, NSFC-12075060); Czech Republic: PRIMUS Research Programme (PRIMUS/21/SCI/017); European Union: European Research Council (ERC - 948254, ERC 101089007), Horizon 2020 Framework Programme (MUCCA - CHIST-ERA-19-XAI-00), European Union, Future Artificial Intelligence Research (FAIR-NextGenerationEU PE00000013), Italian Center for High Performance Computing, Big Data and Quantum Computing (ICSC, NextGenerationEU); France: Agence Nationale de la Recherche (ANR-20-CE31-0013, ANR-21-CE31-0013, ANR-21-CE31-0022), Investissements d'Avenir Labex (ANR-11-LABX-0012); Germany: Baden-Württemberg Stiftung (BW Stiftung-Postdoc Eliteprogramme), Deutsche Forschungsgemeinschaft (DFG - 469666862, DFG - CR 312/5-1); Italy: Istituto Nazionale di Fisica Nucleare (ICSC, NextGenerationEU); Japan: Japan Society for the Promotion of Science (JSPS KAKENHI JP21H05085, JSPS KAKENHI JP22H01227, JSPS KAKENHI JP22H04944, JSPS KAKENHI JP22KK0227); Netherlands: Netherlands Organisation for Scientific Research (NWO Veni 2020 - VI.Veni.202.179); Norway: Research Council of Norway (RCN-314472); Poland: Polish National Agency for Academic Exchange (PPN/PPO/2020/1/00002/U/00001), Polish National Science Centre (NCN 2021/42/E/ST2/00350, NCN OPUS nr 2022/47/B/ST2/03059, NCN UMO-2019/34/E/ST2/00393, UMO-2020/37/B/ST2/01043, UMO-2021/40/C/ST2/00187, UMO-2022/47/O/ST2/00148); Slovenia: Slovenian Research Agency (ARIS grant J1-3010); Spain: BBVA Foundation (LEO22-1-603), Generalitat Valenciana (Artemisa, FEDER, IDIFEDER/2018/048), La Caixa Banking Foundation (LCF/BQ/PI20/11760025), Ministry of Science and Innovation (MCIN \& NextGenEU PCI2022-135018-2, MICIN \& FEDER PID2021-125273NB, RYC2019-028510-I, RYC2020-030254-I, RYC2021-031273-I, RYC2022-038164-I), PROMETEO and GenT Programmes Generalitat Valenciana (CIDEGENT/2019/023, CIDEGENT/2019/027); Sweden: Swedish Research Council (VR 2018-00482, VR 2022-03845, VR 2022-04683, VR grant 2021-03651), Knut and Alice Wallenberg Foundation (KAW 2017.0100, KAW 2018.0157, KAW 2018.0458, KAW 2019.0447); Switzerland: Swiss National Science Foundation (SNSF - PCEFP2\_194658); United Kingdom: Leverhulme Trust (Leverhulme Trust RPG-2020-004); United States of America: U.S. Department of Energy (ECA DE-AC02-76SF00515), Neubauer Family Foundation.


\clearpage

\printbibliography
 
\clearpage
\input{atlas_authlist}

\end{document}

%% file: atlas_authlist.tex
 
\begin{flushleft}
\hypersetup{urlcolor=black}
{\Large The ATLAS Collaboration}

\bigskip

\AtlasOrcid[0000-0002-6665-4934]{G.~Aad}$^\textrm{\scriptsize 102}$,
\AtlasOrcid[0000-0001-7616-1554]{E.~Aakvaag}$^\textrm{\scriptsize 16}$,
\AtlasOrcid[0000-0002-5888-2734]{B.~Abbott}$^\textrm{\scriptsize 120}$,
\AtlasOrcid[0000-0002-1002-1652]{K.~Abeling}$^\textrm{\scriptsize 55}$,
\AtlasOrcid[0000-0001-5763-2760]{N.J.~Abicht}$^\textrm{\scriptsize 49}$,
\AtlasOrcid[0000-0002-8496-9294]{S.H.~Abidi}$^\textrm{\scriptsize 29}$,
\AtlasOrcid[0000-0002-9987-2292]{A.~Aboulhorma}$^\textrm{\scriptsize 35e}$,
\AtlasOrcid[0000-0001-5329-6640]{H.~Abramowicz}$^\textrm{\scriptsize 151}$,
\AtlasOrcid[0000-0002-1599-2896]{H.~Abreu}$^\textrm{\scriptsize 150}$,
\AtlasOrcid[0000-0003-0403-3697]{Y.~Abulaiti}$^\textrm{\scriptsize 117}$
% ...
\bigskip
\\

% ...
$^{16}$Department for Physics and Technology, University of Bergen, Bergen; Norway.\\
% ...
$^{29}$Physics Department, Brookhaven National Laboratory, Upton NY; United States of America.\\
% ...
$^{35}$$^{(a)}$Facult\'e des Sciences Ain Chock, R\'eseau Universitaire de Physique des Hautes Energies - Universit\'e Hassan II, Casablanca;$^{(b)}$Facult\'{e} des Sciences, Universit\'{e} Ibn-Tofail, K\'{e}nitra;$^{(c)}$Facult\'e des Sciences Semlalia, Universit\'e Cadi Ayyad, LPHEA-Marrakech;$^{(d)}$LPMR, Facult\'e des Sciences, Universit\'e Mohamed Premier, Oujda;$^{(e)}$Facult\'e des sciences, Universit\'e Mohammed V, Rabat;$^{(f)}$Institute of Applied Physics, Mohammed VI Polytechnic University, Ben Guerir; Morocco.\\
% ...
$^{49}$Fakult\"{a}t Physik , Technische Universit{\"a}t Dortmund, Dortmund; Germany.\\
% ...
$^{55}$II. Physikalisches Institut, Georg-August-Universit\"{a}t G\"ottingen, G\"ottingen; Germany.\\
% ...
$^{102}$CPPM, Aix-Marseille Universit\'e, CNRS/IN2P3, Marseille; France.\\
% ...
$^{117}$Department of Physics, New York University, New York NY; United States of America.\\
% ...
$^{120}$Homer L. Dodge Department of Physics and Astronomy, University of Oklahoma, Norman OK; United States of America.\\
% ...
$^{150}$Department of Physics, Technion, Israel Institute of Technology, Haifa; Israel.\\
$^{151}$Raymond and Beverly Sackler School of Physics and Astronomy, Tel Aviv University, Tel Aviv; Israel.\\
% ...
$^{a}$ Also Affiliated with an institute covered by a cooperation agreement with CERN.\\
$^{b}$ Also at An-Najah National University, Nablus; Palestine.\\
$^{c}$ Also at Borough of Manhattan Community College, City University of New York, New York NY; United States of America.\\
$^{d}$ Also at Center for High Energy Physics, Peking University; China.\\
$^{e}$ Also at Center for Interdisciplinary Research and Innovation (CIRI-AUTH), Thessaloniki; Greece.\\
$^{f}$ Also at Centro Studi e Ricerche Enrico Fermi; Italy.\\
% ...

\end{flushleft}
